\documentclass{article}
\usepackage[english]{babel}

\usepackage[letterpaper,top=2cm,bottom=2cm,left=3cm,right=3cm,marginparwidth=1.75cm]{geometry}

\usepackage{mathtools}
\usepackage{amssymb}
\usepackage{amsmath}
\usepackage{amsthm}
\usepackage{svg}
\usepackage{hyperref}
\usepackage{soul}

\newcommand*\diff{\mathop{}\!\mathrm{d}}

\title{Constructing PDFs of spatially dependent fields using finite elements}
\author{Paul M. Mannix, David A. Ham, John Craske}

\begin{document}
\maketitle

\section{Introduction}\label{sec:Intro}

A probability density function (PDF)\footnote{For the sake of using common terminology we refer to 
`probability density functions', noting that our implementation does not assume that such functions always exist and therefore computes `distributions' or `generalised functions'.} of a spatially dependent field provides a means of calculating moments of the field or, equivalently, the proportion of a spatial domain that is mapped to a given set of values. This paper describes a finite element approach to estimating the PDF of a spatially dependent field and its numerical implementation in the Python package \texttt{NumDF}.

Probability distributions emerge naturally in the analysis of a wide range of problems in engineering and the natural sciences. For example, under certain conditions the optimal transport map between two functions of the real line 
is given in terms of their cumulative distribution functions as an increasing rearrangement \cite{VilCboo2009a}. For the same reason, probability distributions underpin the calculation of the available potential energy (APE) of a variable-density flow, which quantifies the proportion of potential energy that is theoretically available for conversion into kinetic energy \cite{lorenz1955available, winters1995available, tseng2001mixing}. APE is important as it can be used to quantify the mixing efficiency of different types of turbulence \cite{peltier2003mixing}, as well as helping to understand the dynamics of atmospheric and oceanic energy cycles \cite{tailleux2013available}.

The common approach to estimating the PDF of a spatially dependent field, obtained via numerical simulation, relies on evaluating the field at a discrete set of points on a equally spaced grid. This grid data is then exploited using standard statistical methods, either by binning the data to obtain a piecewise constant histogram or applying a kernel density estimator to obtain a smooth and continuous function \cite{scott2015multivariate}. These approaches have the advantage that they are easy to apply, scale well to higher dimensional data \cite{Odland2018}, and have well established convergence estimates \cite{scott2015multivariate}. However, when calculating the PDF of continuous functions rather than collections of discrete data points, these methods are not necessarily optimal because they do not make use of all of the information we have at our disposal. As pseudo-spectral, finite element and high order spectral methods, all represent fields in terms of a set of spatially dependent basis functions, a function space approach provides a means of better exploiting their output.


\section{A finite element approach}\label{sec:Method}

Given a function $Y(x)$ over a domain $\Omega_X \ni x$, we want to calculate its corresponding cumulative density function (CDF) $F_Y(y)$ and PDF $f_Y(y) = \diff F_Y(y)/ \diff y$ such that we exploit the structure of the available functions. The CDF of $Y(x)$, which is more regular than its PDF, can be computed as
\begin{equation}
    F_Y(y) = \int_{\Omega_X} \mathbb{I}_{Y(x)<y}(x) \, \diff x, \quad \text{where} \quad \mathbb{I}_{Y(x)<y}(x) =
    \begin{cases}
    1, \quad & Y(x) < y, \\
    0, \quad & \text{otherwise},
    \end{cases}
    \label{eq:CDF}
\end{equation}
denotes the indicator function induced by $Y(x)$. Because the integral of the indicator function is difficult to calculate, a common approach is to estimate the CDF from a set of grid point evaluations of $Y(x)$, which entails information loss. Instead, if we retain information from $Y(x)$ and define the constant extensions $\hat{Y}(x,y) = Y(x), \, \hat{F}_Y(x,y) = F_Y(y)$ over $\Omega_X \times \Omega_Y$, we can directly project $F_{Y}(y)$ into a subspace of $L^2(\Omega_Y)$.

\begin{figure}[t]
    \centering
    \def\svgwidth{0.995\textwidth}
    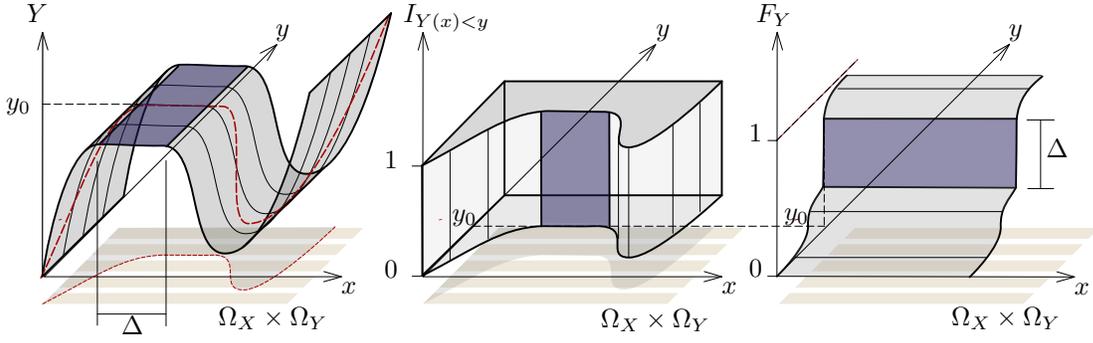
    \caption{The CDF $F_Y$ of $Y(x)$ is constructed by projecting the integral of the indicator function \eqref{eq:CDF} onto the  discontinuous piecewise linear finite element space (DG1), which in this schematic has 5 elements.  (left) The extended function $\hat{Y}(x,y) = Y(x)$, (middle) the indicator function $\mathbb{I}_{Y(x) < y}$ induced by $Y(x)$ and (right) the extended CDF $\hat{F}_Y(x,y) = F_Y(y)$ are defined on the domain $\Omega_X \times \Omega_Y$ which is discretised in $y$ but not in $x$. By representing $F_Y$ using DG1, regions of width $\Delta$ where $Y(x) = y_0$ is constant are correctly described as jumps of height $\Delta$ at $y_0$ in the CDF.}
    \label{fig:Schematic}
\end{figure}

Choosing the space of discontinuous piecewise linear functions (also known as DG1) for the CDF, we define the finite element basis $V_{\mathsf{F}_Y} = \text{span}\{\phi_i(y)\}$ such that $\mathsf{F}_{Y}(y) = \sum_i \mathsf{F}_i \phi_i(y)$ represents the finite element approximation of $F_Y$. The projection of $\mathsf{F}_Y$ into $V_{\mathsf{F}_Y}$ is then given by
\begin{equation}
    \int_{\Omega_X \times \Omega_y} \hat{\mathsf{F}}_Y(x,y) \hat{v}(x,y) \, \diff y  \diff x = \int_{\Omega_X \times \Omega_Y} \mathbb{I}_{Y(x) < y}(x) \hat{v}(x,y) \, \diff y  \diff x,
\end{equation}
where $\hat{v}(x,y) = v(y), \, \hat{\mathsf{F}}_Y(x,y) = \mathsf{F}_Y(y)$ denotes the extension of $v,\mathsf{F}_Y \in V_{\mathsf{F}_Y}$ over $\Omega_X \times \Omega_Y$. Substituting for $\hat{\mathsf{F}}_Y$, the component-wise projection is
\begin{equation}
    \sum_{i,j} \left[ \int_{\Omega_Y} \phi_i(y) \phi_j(y) \, \diff y  \right] \mathsf{F}_j = \sum_i \left[ \int_{\Omega_X \times \Omega_Y} \mathbb{I}_{Y(x) < y}(x) \hat{\phi}_i(y, x) \, \diff y  \diff x \right],
    \label{eq:Proj_CDF}
\end{equation}
where $\hat{\phi}_i(x,y) = \phi_i(y)$ denotes the extension of $\phi_i$ over $\Omega_X \times \Omega_Y$. The extension of $Y$ and $F_Y$ is shown schematically in figure \ref{fig:Schematic}, along with the discretisation of $\Omega_{X} \times \Omega_Y$ into elements. By extending the domain we ensure that the expression for $\mathbb{I}_{Y(x) < y}(x)$ in the right hand side of \eqref{eq:Proj_CDF} remains exact, which  guarantees that we calculate the finite element approximation of the exact CDF. If instead we first integrated the indicator function with respect to $x$ and then projected the resulting numerical approximation of \eqref{eq:CDF} into a finite element space, we would be numerically approximating the CDF at a set of discrete grid points prior to its projection into a finite element space. The projection defined by \eqref{eq:Proj_CDF} avoids this additional level of approximation but at the cost of increasing the dimensionality of the calculation. For example, if $\Omega_X$ is the $n$ dimensional space $\mathbb{R}^n$, the function $Y(x)$ will, when extended over $\Omega_X \times \Omega_Y$, result in a $n+1$ dimensional problem.

As the projection into $V_{\mathsf{F}_Y}$ does not guarantee a non-decreasing CDF, it is necessary to apply a slope limiter after evaluating \eqref{eq:Proj_CDF}. As we have chosen the function space DG1, guaranteeing a non-decreasing $\mathsf{F}_Y(y)$ amounts to ensuring a non-negative slope within each element and a non-negative jump between successive elements. Having chosen degrees of freedom (dofs) located at the element boundaries, ensuring a non-decreasing CDF is equivalent to ensuring monotonically non-decreasing dofs with respect to $y$. If a discontinuous piecewise quadratic or higher order function space was chosen, the slope limiter would be more complicated than ensuring monotonically non-decreasing dofs. An additional motivation for restricting our implementation to DG1 is that the integrand on the right hand side of \eqref{eq:Proj_CDF} (as shown schematically in the middle frame of figure \ref{fig:Schematic}) is discontinuous. Given that the numerical quadrature employed will not be exact, it is therefore unlikely that any benefit would be gained by using higher order elements.

The PDF $f_Y(y)$ is the derivative of the CDF $F_Y(y)$, but as the numerical approximation $\mathsf{F}_Y(y)$ is discontinuous between elements, its derivative lacks the regularity required to be expressed as a function. This leads to the definition of the finite element approximation of the PDF as the distribution
\begin{equation}
\mathsf{f}_Y[v] \coloneqq \int_{\Omega_Y} v(y) \, \diff \mathsf{F}_Y (y), \quad \text{where} \quad v \in C_0(\Omega_Y).
\label{eq:PDF_definition}
\end{equation}
Although the PDF cannot be expressed as a function, Lebesgue's decomposition theorem \cite{rudin1987real} (and the finite dimensional setting of this problem) allows this distribution to be written as a sum of two continuous linear functionals
\begin{equation}
\mathsf{f}_Y[v] = \langle v, \mathsf{f}_0 \rangle_{V_0} + \langle v, \mathsf{f}_1 \rangle_{V_1},
\end{equation}
over the Hilbert spaces $V_0$ and $V_1$ associated with the continuous part of the PDF within the elements, and the singular contribution to the PDF from jumps at the facets $\Gamma$, which we define as the set of interior boundaries between elements. This is useful because although we cannot calculate the functional form of the PDF, we can compute its Riesz-representation as the functions $\mathsf{f}_0(y) \in L^2(\Omega_Y)$ and $\mathsf{f}_1(y) \in L^2(\Gamma)$.

As a consequence of \eqref{eq:PDF_definition} $\mathsf{f}_0$ lies within the space of discontinuous piecewise constant functions and we define the basis $V_0 = \text{span}\{\varphi_i(y)\}$ such that $\mathsf{f}_0(y) = \sum_i \mathsf{f}^i_0 \varphi_i(y)$. As element boundaries in $L^2(\Omega_Y)$ have measure zero the projection over all elements is 
\begin{equation}
    \int_{\Omega_Y} v(y) \mathsf{f}_0(y) \, \diff y = \int_{\Omega_Y} v(y) \partial_y \mathsf{F}_Y(y) \, \diff y, \quad \text{where} \quad v \in V_0.
\end{equation}
At a facet $y_e \in \Gamma$ the distributional derivative is therefore given, for a suitable test function $\varphi(y)$, by $-[[\mathsf{F}_Y(y_e)]] \varphi(y_e)$, where $[[\mathsf{F}_Y(y_e)]]$ is the jump in the CDF. Choosing a space of continuous functions with nodes at the cell boundaries and defining the basis $V_1 = \text{span}\{\varphi_i(y)\}$ such that $\mathsf{f}_1(y) = \sum_i \mathsf{f}^i_1 \varphi_i(y)$, we define the projection over the set of facets as
\begin{equation}
    \mathsf{f}^i_1 = - \sum_{y_e \in \Gamma} \varphi_i(y_e) [[\mathsf{F}_Y(y_e)]], \quad \text{where} \quad \varphi_i \in V_1.
\end{equation}
If the true CDF $F_Y$ is smooth and continuous, the jumps in its finite element approximation $[[\mathsf{F}_Y(y_e)]]$ at the facets will tend to zero as the approximation converges. On the other hand should the true CDF contain a step corresponding, for example, to a region of $\Omega_X$ where $Y(x)$ is constant (as indicated in figure \ref{fig:Schematic}), the jump $[[\mathsf{F}_Y(y_e)]]$ will be correctly associated with a Dirac measure. 

\section{Calculating the available potential energy (APE)}\label{sec:Examples}

The volume-averaged APE of a buoyancy field $B(\boldsymbol{x})$ with vertical coordinate $Z(\boldsymbol{x})$ is given by the difference of its total and (minimal) background potential energy according to 
\begin{equation}
\mathbb{E}[\beta^*Z]  - \mathbb{E}[BZ] = \int_{\Omega_Z} \beta^*(z) z \, \diff F_Z(z) - \frac{1}{V} \int_{\Omega_X} B(\boldsymbol{x}) Z(\boldsymbol{x}) \, \diff \boldsymbol{x},
\end{equation}
where $V$ is the volume of the domain and the reference buoyancy $\beta^*(z) = F^{-1}_B \circ F_Z(z)$, 
corresponds to an adiabatic volume preserving rearrangement of the fluid into a state that minimizes its total potential energy. Calculating APE using $\beta^{*}$ therefore requires a means of calculating the inverse of a CDF. 

\begin{figure}[h!]
    \centering
    \includegraphics[scale=0.5]{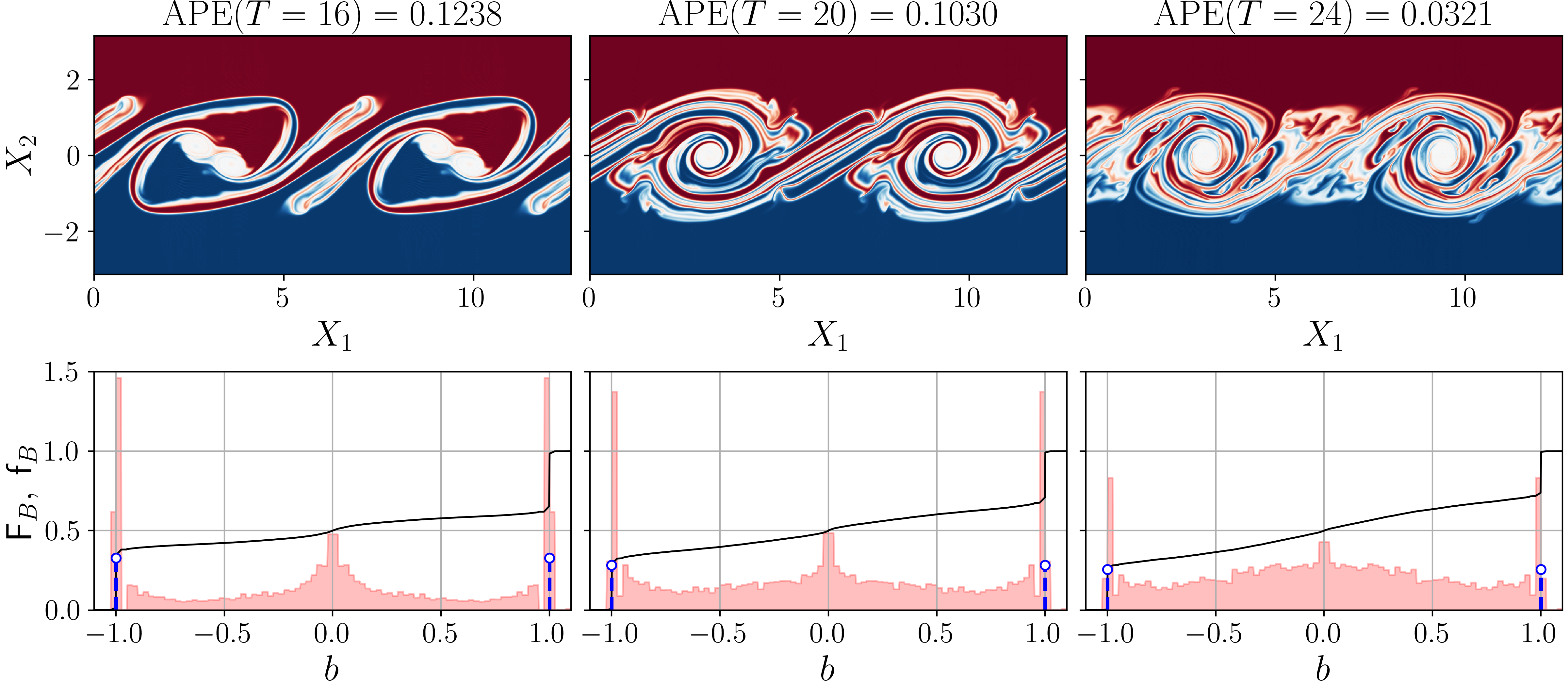}
    \caption{(Top) As the two dimensional Kelvin-Helmholtz instability evolves, the stably stratified interface of negative/positive buoyancy (blue/red) is deformed. This process first creates APE by displacing negatively buoyant fluid upward and positively buoyant fluid downward. As the Kelvin-Helmholtz billows break kinetic energy increases and APE is destroyed. (Bottom) The finite element approximation of the PDF of buoyancy in terms of its continuous $\mathsf{f}_0$ (filled red) and singular $\mathsf{f}_1$ (dashed blue) measures becomes more uniform as time evolves and APE reduces, while the finite element approximation of the CDF (solid black) which initially approximates a step function becomes more tilted.}
    \label{fig:KH}
\end{figure}

Our approach exploits the fact that our finite element approximation of $F_B(b)$ which we call $\mathsf{F}_B(b)$, is a discontinuous piecewise linear function which can be inverted element-wise. We use the dofs of $\mathsf{F}_B(b)$ which are located at the element boundaries to specify a non-uniform mesh $\Omega_p \ni p$ on which the continuous piecewise linear function $\mathsf{F}^{-1}_B(p)$ is defined. Then by locating the dofs of the CDF at the element boundaries and using this non-uniform mesh, the inverse CDF is trivially computed by assigning the dofs of $\mathsf{F}^{-1}_B(p)$, the $b$ values corresponding to the dofs of $\mathsf{F}_B(b)$. Although APE is typically computed using the reference height $Z^*(b) = F^{-1}_Z \circ F_B(b)$ (see for example \cite{peltier2003mixing}), we have instead chosen to use $\beta^*(z)$. This is because using $Z^*(b)$ becomes problematic when $F_B(b)$ is multivalued. For example at $Y(x) = y_0$ in figure \ref{fig:Schematic} or $b = \pm 1$ in figure \ref{fig:KH}, one needs to specify how to evaluate $F_Y(y)$ or $F_B(b)$ at these points in order to calculating the APE using $Z^*(b)$. As the finite element approximation of the CDF $\mathsf{F}_B(b)$, is easily inverted, and because $\mathsf{F}^{-1}_B(p)$ is a well defined continuous function, no such choice is needed for our approach. 

Figure \ref{fig:KH} illustrates application of the method to calculate the APE of a time evolving two-dimensional Kelvin-Helmholtz instability computed using the pseudo-spectral code Dedalus \cite{burns2020dedalus}. The Chebyshev-Fourier basis employed uses $1024$ modes in the horizontal, $512$ polynomials in the vertical and returns a set of non-uniformly spaced collocation points. Interpolating these points using the basis functions, \texttt{NumDF} can directly evaluate the CDF and its inverse with an $L^2$ error which scales as $\mathcal{O}(N_y^{-2})$ and the PDF with an $L^1$ error which scales as $\mathcal{O}(N_y^{-1/2})$ for a sufficiently large quadrature degree, where $N_y$ is the number of finite-elements. In contrast to estimating the PDF using a histogram, whose $L^1$ error, subject to choosing the optimal bin-width/bandwidth, scales as $\mathcal{O}(n^{-1/3})$ and $\mathcal{O}(n^{-2/5})$ respectively where $n$ is number data points \cite{devroye1985nonparametric}, no free parameters must be optimised in the finite-element approach.

Although the current implementation only treats one and two dimensional functions we envisage its extension to three dimensions for further work. Similarly, although we have only considered rectangular geometries in this paper, future work can seek to extend to non-standard domains, where the question of how to sample data to apply standard statistical methods is often unclear thus potentially making our method particularly useful. The complete documentation for \texttt{NumDF} is available on the NumDF website \footnote{NumDF website - https://mannixp.github.io/D.stratify-pdfe/}, while all scripts and data used to generate the figures and conduct the convergence tests referred to in the previous paragraph are available on Zenodo \cite{mannix_2024_14259619}. 

\section*{Acknowledgements}
This work was supported by the Engineering and Physical Sciences Research Council [grant number EP/V033883/1] as part of the [D$^{*}$]stratify project. The authors would also like to thank the Firedrake project \cite{FiredrakeUserManual} for making the code development in this paper possible.

\bibliographystyle{plain}
\bibliography{sample}

\begin{thebibliography}{10}

\bibitem{burns2020dedalus}
Keaton~J Burns, Geoffrey~M Vasil, Jeffrey~S Oishi, Daniel Lecoanet, and Benjamin~P Brown.
\newblock Dedalus: A flexible framework for numerical simulations with spectral methods.
\newblock {\em Physical Review Research}, 2(2):023068, 2020.

\bibitem{devroye1985nonparametric}
Luc Devroye and László Györfi.
\newblock {\em Nonparametric density estimation: The $L_1$ View}.
\newblock John Wiley, 1985.

\bibitem{FiredrakeUserManual}
David~A. Ham, Paul H.~J. Kelly, Lawrence Mitchell, Colin~J. Cotter, Robert~C. Kirby, Koki Sagiyama, Nacime Bouziani, Sophia Vorderwuelbecke, Thomas~J. Gregory, Jack Betteridge, Daniel~R. Shapero, Reuben~W. Nixon-Hill, Connor~J. Ward, Patrick~E. Farrell, Pablo~D. Brubeck, India Marsden, Thomas~H. Gibson, Miklós Homolya, Tianjiao Sun, Andrew T.~T. McRae, Fabio Luporini, Alastair Gregory, Michael Lange, Simon~W. Funke, Florian Rathgeber, Gheorghe-Teodor Bercea, and Graham~R. Markall.
\newblock {\em Firedrake User Manual}.
\newblock Imperial College London and University of Oxford and Baylor University and University of Washington, first edition edition, 5 2023.

\bibitem{lorenz1955available}
Edward~N Lorenz.
\newblock Available potential energy and the maintenance of the general circulation.
\newblock {\em Tellus}, 7(2):157--167, 1955.

\bibitem{mannix_2024_14259619}
Paul Mannix, David Ham, and John Craske.
\newblock {Supporting code and data for the paper: Estimating the PDF of a spatially dependent field using a finite element approach}, December 2024.

\bibitem{Odland2018}
Tommy Odland.
\newblock Kdepy: Kernel density estimation in python (v0.9.10).
\newblock {\em Zenodo. https://doi.org/10.5281/zenodo.2392268}, 2018.

\bibitem{peltier2003mixing}
WR~Peltier and CP~Caulfield.
\newblock Mixing efficiency in stratified shear flows.
\newblock {\em Annual Review of Fluid Mechanics}, 35(1):135--167, 2003.

\bibitem{rudin1987real}
W.~Rudin.
\newblock {\em Real and Complex Analysis}.
\newblock Mathematics series. McGraw-Hill, 1987.

\bibitem{scott2015multivariate}
David~W Scott.
\newblock {\em Multivariate density estimation: theory, practice, and visualization}.
\newblock John Wiley \& Sons, 2015.

\bibitem{tailleux2013available}
R{\'e}mi Tailleux.
\newblock Available potential energy and exergy in stratified fluids.
\newblock {\em Annual Review of Fluid Mechanics}, 45(1):35--58, 2013.

\bibitem{tseng2001mixing}
Yu-heng Tseng and Joel~H Ferziger.
\newblock Mixing and available potential energy in stratified flows.
\newblock {\em Physics of Fluids}, 13(5):1281--1293, 2001.

\bibitem{VilCboo2009a}
Cédric. Villani.
\newblock {\em Optimal Transport Old and New}.
\newblock Grundlehren der mathematischen Wissenschaften, A Series of Comprehensive Studies in Mathematics, 338. Springer Berlin Heidelberg, Berlin, Heidelberg, 1st ed. 2009. edition, 2009.

\bibitem{winters1995available}
Kraig~B Winters, Peter~N Lombard, James~J Riley, and Eric~A D'Asaro.
\newblock Available potential energy and mixing in density-stratified fluids.
\newblock {\em Journal of Fluid Mechanics}, 289:115--128, 1995.

\end{thebibliography}

\end{document}